\documentclass[prd,twocolumn, reprint,amsmath,amssymb, aps, superscriptaddress]{revtex4}

\usepackage{graphicx}
\usepackage{epsfig}
\usepackage{dcolumn}
\usepackage{bm}

\begin{document}

\title{Testing the Newborn Pulsar Origin of \\Ultrahigh Energy Cosmic Rays with EeV Neutrinos}

\author{Ke Fang}
\affiliation{Department of Astronomy \& Astrophysics, Kavli Institute for Cosmological Physics, The
  University of Chicago, Chicago, Illinois 60637, USA.}
\author{Kumiko Kotera}
\affiliation{Institut d'Astrophysique de Paris, UMR 7095 - CNRS, Universit\'e Pierre $\&$ Marie Curie, 98 bis boulevard Arago, 75014, Paris, France}
\author{Kohta Murase}
\affiliation{Hubble Fellow -- Institute for Advanced Study, Princeton, NJ 08540, USA}
\author{Angela V. Olinto}
\affiliation{Department of Astronomy \& Astrophysics, Kavli Institute for Cosmological Physics, The
  University of Chicago, Chicago, Illinois 60637, USA.}

\date{\today}

\begin{abstract}
Fast-spinning newborn pulsars are intriguing candidate sources of ultrahigh energy cosmic rays (UHECRs). The acceleration of particles with a given composition in a fraction of the extragalactic pulsar population can give a consistent explanation for the measurements of the Auger Observatory. We calculate the associated diffuse neutrino flux produced while particles cross the supernova ejecta surrounding the stars. We show that in the minimal pulsar scenario that are compatible with the UHECR data, the effective optical depth to hadronuclear interactions is larger than unity at ultrahigh energies. Thus, even in the most pessimistic case, one expects energy fluxes of $\sim 0.1 - 1 \,\rm EeV$ neutrinos that should be detectable with IceCube or Askaryan Radio Array within a decade.
\begin{description}
\item[PACS numbers]{95.85.Ry, 97.60.Jd, 98.70.Sa}
\end{description}
\end{abstract}

\maketitle

\section{Introduction}
The two most popular candidate sources of ultrahigh energy cosmic rays (UHECRs): active galactic nuclei (AGN) and gamma-ray bursts (GRBs) are challenged by the latest cosmic-ray measurements by the Pierre Auger Observatory. 
In particular, AGN are not favored as nuclei-rich sources \cite{Horiuchi12, Lemoine:2009pw}, given that the latest Auger observations seem to indicated an increasing mass composition at the highest energies \cite{AugerIcrc11}.  The GRB population is viable but barely meets the energetics and spectral criteria to fit the observations (e.g., Refs.~\cite{KO11,Murase08}). 

Newborn pulsars on the other hand could satisfy these criteria, given their metal-rich surfaces, high number density, and huge energetics, especially for those spinning close to millisecond periods at birth \cite{Venkatesan97}. 
Heavy ions could be seeded into the current sheets of the neutron star wind \cite{Hoshino92,Gallant94,Arons03} and get bulk acceleration by energy conversion of the wind Poynting flux into kinetic energy like a unipolar inductor.  Particle acceleration could happen somewhere before or at the termination shock, far away from the light cylinder, to avoid radiative losses \cite{Arons03,Blasi00,Murase09,FKO12,FKO13}.  The crossing of the surrounding supernova ejecta tends to prevent the escape of particles at the earliest times. As the ejecta expands and becomes thinner, the heaviest nuclei, accelerated to higher energies than lighter ones because of their charge, are able to escape at energies $E>10^{20}\,$eV. Nuclei interactions with the baryonic and radiative backgrounds of the ejecta produce secondary nucleons that soften the overall cosmic-ray spectrum. Interestingly, these interactions also lead to the production of EeV neutrinos (see Ref.~\cite{Murase09} for the magnetar scenario).

After propagation in the intergalactic medium, and integrating over a fraction of the whole extragalactic newborn pulsar population, it is then possible to explain the spectrum, composition, and anisotropy measurements of the Auger Observatory consistently.  Moreover, Galactic pulsar counterparts can account for the flux of cosmic rays in the region below the ankle, and bridge the gap between a component due to acceleration in Galactic supernova remnants, and extragalactic sources \cite{FKO13}. 

In this work, we show that within the parameter-space allowed by this newborn pulsar scenario to reproduce the observed cosmic-ray data, $\sim0.1-1$~EeV neutrino production occurs {\it efficiently} and the diffuse neutrino flux is detectable by IceCube, KM3Net, Askaryan Radio Array (ARA), and the Antarctic Ross Ice-Shelf ANtenna Neutrino Array (ARIANNA) within a decade even in the most pessimistic case.  
Our estimates lie sensibly above the IceCube-5 years sensitivity in the $10^{18}\,$eV-energy range, and are below the current IceCube sensitivity. This is a crucial test, since nondetections can rule out the minimal newborn pulsar scenario for UHECRs within the next decade.  
Testing the newborn pulsar scenario is intriguing, especially if the heavy composition of UHECRs is confirmed.  Photohadronic neutrinos from UHECR sources such as GRBs and AGN are difficult to detect if UHECRs are dominantly nuclei \cite{Murase08}.

We first introduce the model of UHECR and associated neutrino production for a single pulsar. We then present our results integrated for populations of sources with parameters that fit the Auger measurements including both the spectrum and composition. We finally discuss the robustness of our diffuse neutrino flux estimate and weigh the power of this test to probe the newborn pulsar origin of UHECRs.

\section{Neutrinos from a single pulsar}
A newborn pulsar with initial spin period $P_{\rm i}=1\,{\rm ms} \,P_{\rm i, -3}$, surface magnetic field $B=10^{13}\,{\rm G}\,B_{13}$ and radius $R=10\,\rm km$, spins down due to electromagnetic losses over a typical timescale $\tau_{\rm EM}=1{\,\rm yr}\,B_{13}^{-2}\,P_{\rm i,-3}^{2}$.  Assuming that particles of charge $Z$ can recuperate a fraction $\eta=0.1\,\eta_{-1}$ of the wind Poynting flux at time $t$, particles each gain energy  $
E_{\rm CR} (t) = 7\times 10^{18} \,{\rm eV}\,\eta_{-1}\,Z\,B_{13}\,P_{\rm i, -3}^{-2}\,\left(1+{t}/{\tau_{\rm EM}}\right)^{-1}$ \cite{Blasi00,Arons03}.  
{ In the standard pulsar model, the wind is dominated by pairs outside the equatorial current sheet, and its loading rate is much larger than the Goldreich-Julian rate \cite{Goldreich69}.  But the return current may largely consist of ions, where the ion injection rate around the equatorial sector is comparable to the Goldreich-Julian rate \cite{Hoshino92,Gallant94,Arons03}. The deviation can be accounted for in the prefactor $f_{\rm s}<1$ (see Eq.~\ref{eq:diff_nu}). 

In our minimal pulsar scenario, the cosmic ray spectrum injected during the pulsar spin-down is \citep{Blasi00, Arons03}: ${{\rm d}N_{\rm CR}}/{{\rm d}E}=9{c^2\,I}/({8Z\,e\,\mu})\,E^{-1}$, where $I=10^{45}\,I_{45}\,\rm g\,cm^2$ is the moment of inertia of the star.  If the stochastic acceleration mechanism like the Fermi mechanism is additionally involved, this injection spectrum can be modified and softened to a power-law closer to $\propto E^{-2}$.  The impact of such modifications are discussed at the end of this paper, and our conclusion does not change.  

Particle acceleration in the wind would occur far away from the light cylinder to avoid losses due to the radiation from the cooling stellar envelope that would be heated by emission from the wind bubble and radioactive nuclei.  It has been shown that nuclei can survive from photodisintegration losses if radiation fields are thermal \cite{FKO12}.  Synchrotron emission from electrons accelerated at the termination shock can destroy nuclei~\cite{mdt14} but details are highly uncertain.  We assume that non-thermal radiation fields allow nucleus-survival, which may be the case if the termination shock is still hydrodynamically weak and the wind is Poynting-dominated \cite{2005ApJ...628..315Z}.}

\begin{figure}[t]
\centering
\epsfig{file=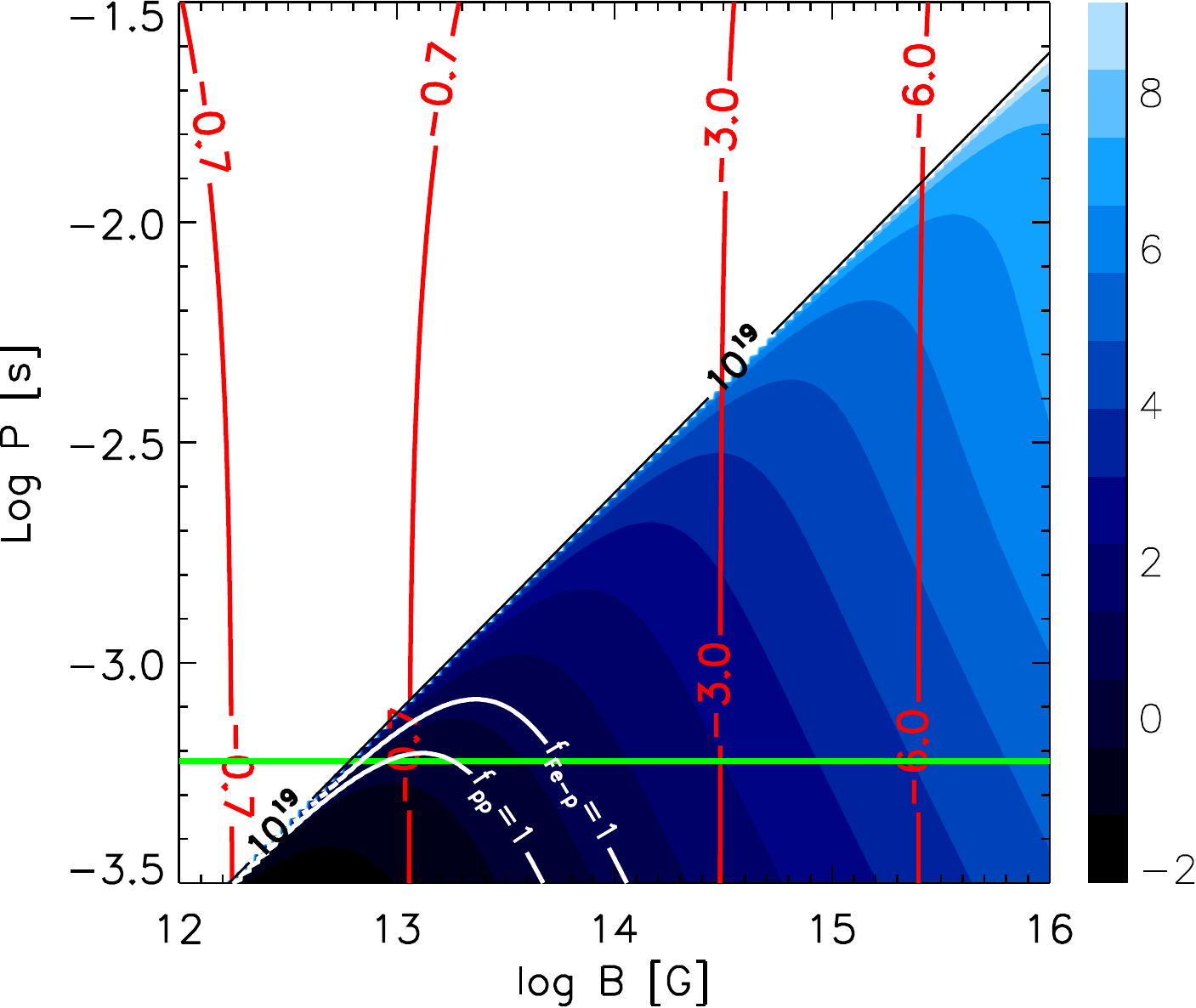,width=0.85\columnwidth,clip=}
\caption{\label{fig:opacity} {Effective optical depth  $f_{pp}$ of hadron interactions in a $10\,M_\odot$ supernova ejecta at the time a pulsar with initial period $P$ and surface magnetic field $B$ accelerates $10^{19}\,\rm eV$ protons (regardless of energy losses). Pulsars that are capable of accelerating protons to above $10^{19}\,\rm eV$ lie under the black line ($\eta=0.1$ assumed). The blue shaded contours span from $f_{pp}= 10^{-2}$ to $10^{9}$ (among which $f_{pp} =1$ is indicated in white). $f_{\rm Fe-p}=1$ is also shown for comparison.
The overplotted red lines indicate the probability distribution function $f(P, B) = f(P) \, f(B)$ of the pulsars. We
assume that all pulsars have initial spin periods above the green line, which indicates the minimum spin period of a stable neutron ￼￼star $P_{\rm min} \approx 0.6\,\rm ms $ \cite{Haensel99}.
Both $f_{pp}$ and $f(P, B)$ are in logarithmic scale.}
}
\end{figure}

Accelerated particles then travel through the expanding supernova ejecta surrounding the star. The ejecta is modeled as a shell spherically expanding at velocity $\beta=({2E_{\rm ej}}/{M_{\rm ej}\,c^2})^{1/2}$ and with column density  $y_{\rm SN}(t) = \int\,\rho_{\rm SN}\,{\rm d}R_{\rm SN} \sim 2\,M_{\rm ej,1}^2E_{\rm ej,52}^{-1} t_{\rm yr}^{-2}~{\rm g\,cm}^{-2}$ at one year, with $M_{\rm ej}=10\,M_{\rm ej, 1}\,M_\odot$ is the ejecta mass and $E_{\rm ej}=10^{52}\,E_{\rm ej, 52}\,{\rm erg}=E_{\rm rot}+E_{\rm exp}$ is the ejecta energy that includes both the star's rotational and the supernova explosion energy \citep{FKO12}.  Note that $\beta\sim 0.03$ for a hypernova with $E_{\rm ej}\sim10^{52}\,\rm erg$ and $\beta\approx 0.01$ for a ``typical'' Type II supernova with $E_{\rm ej}\sim 10^{51}\,\rm erg$.
{ The magnetic field in the ejecta is negligible at that time due to the adiabatic expansion. We may assume a uniform ambient density over the shell at any given $t$, although more detailed density evolution of the ejecta depends on supernova types, and $y_{\rm SN}(t)$ provides a good estimate for the evolution of the integrated column density 
\cite{FKO12}.  
At $t=1{\,\rm year}\,t_{\rm yr}$, the proton-proton ($pp$) interaction has an effective  optical depth 
$f_{pp}={R_{\rm ej}}\,{n_{p}\sigma_{pp}\kappa}\sim0.2\,M_{\rm ej, 1}\,\beta_{-1.5}^{-2}\,t_{\rm yr}^{-2}$, with ejecta size $R_{\rm ej}(t)=\beta c t$, interaction cross section $\sigma_{pp}\sim100\,\rm mb$, and inelasticity $\kappa\sim0.7$. Note that, since $\sigma_{Np}\sim A^{2/3} \sigma_{pp}$ and $\kappa\sim0.7/A$, the effective optical depth for nuclei ($f_{Np}$) is reduced by $\sim A^{1/3}$. 
At early times when UHECR production is possible, the secondary nuclei, nucleons and pions should efficiently interact with target nucleons and produce higher order nuclei, neutrinos and pions \cite{Murase09}. Pions interact with protons with cross section $\sigma_{\pi p} \sim 5 \times 10^{-26}\,\rm cm^2$, producing additional neutrinos and pions that undergo further $\pi p$ interaction. This cascade continues until ${t_{\pi \rm p}}>{\gamma_\pi\,\tau_\pi}$, when the $\pi p$ interaction time $t_{\pi p}$ becomes longer than the primary or secondary pion's life time $\tau_\pi$ in the lab frame. This critical time is  $t_{\pi}=2\times 10^6 \,{\rm s}\, \eta_{-1}^{1/4}\,M_{\rm ej,1}^{1/4}\,B_{13}^{-1/4}\,\beta_{-1.5}^{-3/4}$ \citep{Murase09}. Then charged pions stop interacting and decay into neutrinos via $\pi^\pm\rightarrow e^\pm+\nu_e(\bar{\nu}_e)+\bar{\nu}_\mu +\nu_\mu$. 

At the time when $Z10^{19}\,\rm eV$ cosmic rays (which correspond to $\sim 5\times10^{17}\,\rm eV$ neutrinos around the peak energy) are accelerated, newborn pulsars are surrounded by ejecta with effective opacity (including energy losses) $f_{pp}\gtrsim1$ and $f_{Np}\gg1$. This leads to the production of secondary nucleons, for which the opacity is $f_{pp}>10$, as shown in Fig.~\ref{fig:opacity}. This figure shows that neutrino production in the parameter space that can produce UHECRs is unavoidable. 
Our results are only mildly sensitive to the ejecta mass as long as $M_{\rm ej}\gtrsim 3M_{\odot}$ \cite{FKO12}.  Thus, for typical Type II supernovae, hadron interactions and the subsequent production of EeV neutrinos should be efficient in this minimal newborn pulsar model.}

In our work the interactions with the baryonic background of the supernova ejecta (assumed to consist of hydrogen, as more sophisticated composition have little effect on the escaped cosmic ray characteristics \cite{FKO12}) were calculated by Monte Carlo for injected nuclei and their cascade products as in Refs.~\cite{KAM09, FKO12, FKO13}. Tables for $\pi p$ interactions were generated using the hadronic model EPOS \cite{Werner06}. Note that neutrinos from secondary nuclei contribute significantly and dominate over leading nuclei in neutrino production.

\section{Diffuse neutrino intensity}

According to Ref.~\cite{Faucher06}, the distribution of pulsar birth spin periods, $f(P)$, is normal, centered at $300$ ms, with standard deviation of $150$ ms. Note that among this population, the sources capable of producing the highest energy cosmic rays are (rare) pulsars born with millisecond periods and average magnetic fields \cite{FKO12}. The initial magnetic field follows a log-normal distribution $f(B)$ with $\langle \log(B/\rm G)\rangle = 12.65$ and $\sigma (\log B) = 0.55$. The averaged neutrino and cosmic ray spectrum from the pulsar population is then \cite{FKO13}
$\langle{dN}/{dE}\rangle=\int\,{dN}/{dE}(P,B)\,f(P)\,dP\,f(B)\,dB$.
This population of extragalactic pulsars is expected to contribute to the diffuse neutrino background, which is given by
\begin{eqnarray}
\Phi_{\nu}&=& \frac{f_{\rm s}}{4\pi}\,\int_0^{z_{\rm max}} \int_0^{t_{\nu}} \frac{{\rm d}N_\nu}{{\rm d}t'\,{\rm d}E_\nu\,4\pi D^2}\,{\rm d}t' \,  4\pi D^2  \label{eqn:diff_nu1} \\ \nonumber
&\times &\dot{\Re}(z)\,\frac{{\rm d}D}{{\rm d}z}\,{\rm d}z  \\ 
&=& \frac{c\,f_{\rm s}}{4\pi} \int_0^{z_{\rm max}}\dot{\Re}(z)\frac{{\rm d}N[E(1+z)]}{dE'}(1+z)\,\frac{dt}{dz}dz \label{eq:diff_nu}
\end{eqnarray}
The inner integral in equation~\ref{eqn:diff_nu1} counts the neutrinos emitted by each pulsar toward the earth during its neutrino-loud lifetime 
$t_\nu=\min{(t_{pp}, \, t_{\pi})}$.  
In simulations, this integral is calculated by summing up the spectra from pulsars with $19 \times 19$ sets of $(P_{\rm i}, \log B)$ over the pulsar distributions.
This averaged contribution from an individual star is then integrated over the entire source population in the universe up to the first stars, corresponding to redshift $z_{\rm D}\approx 11$.  Note   $E'=(1+z) E$ in equation~\ref{eq:diff_nu} is the redshifted energy at the source. 
The local birth rate of pulsars is set to the rate of core-collapse supernova, of order $\dot{\Re}(0)\approx 1.2\times10^{-4}\,\rm yr^{-1}\,Mpc^{-3}$ \citep{1991ARA&A..29..363V}, as a large fraction of such events produce neutron stars \cite{Woosley02}.
The source emissivity is assumed to either follow the star formation rate (SFR) \citep{2008ApJ...683L...5Y}, or be uniform over time.  
The ion injection rate is reduced by the pair loading, particle acceleration mechanisms, and geometry of the current sheet, all of which are taken into account by a prefactor $f_{\rm s} <1$. 
Cosmic rays lose energy during their propagation in the IGM by interactions against cosmic radiation backgrounds, pair production and cosmological expansion. We use here the propagation calculations by Monte-Carlo done in Ref.~\cite{FKO13}. Then $f_{\rm s}$ is obtained by fitting the simulation output to the observations.  {Note that the escaping cosmic-ray flux is also proportional to $f_s$, so the resulting neutrino flux does {\it not} depend on $f_s$ since it is directly normalized by the cosmic-ray data.}  The injected elements are divided into three groups (adding more elements does not refine the fit, and introduces unnecessary free parameters): Hydrogen, Carbon group (CNO), and Iron. The relative abundance of these groups is chosen to best fit the spectrum and the main estimators of the composition measured by Auger, namely the mean air-shower elongation rate $\langle X_{\rm max}\rangle$ and its root mean square ${\rm RMS}(X_{\rm max})$.\\

\begin{figure}[t]
\centering
\epsfig{file=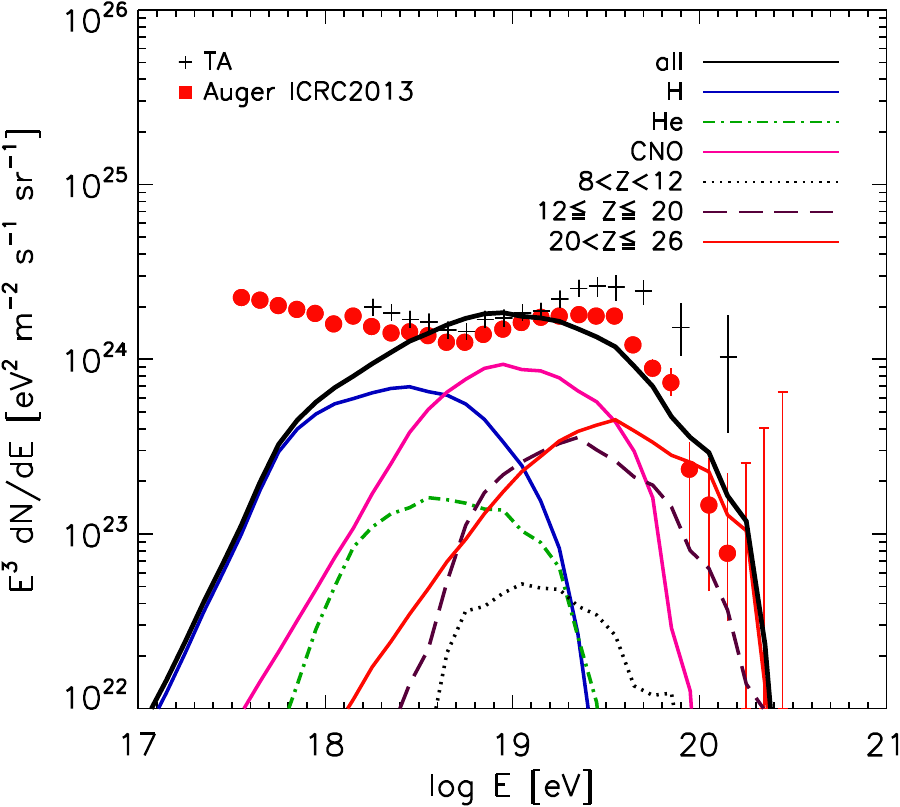,width=0.85\columnwidth,clip=}
\epsfig{file=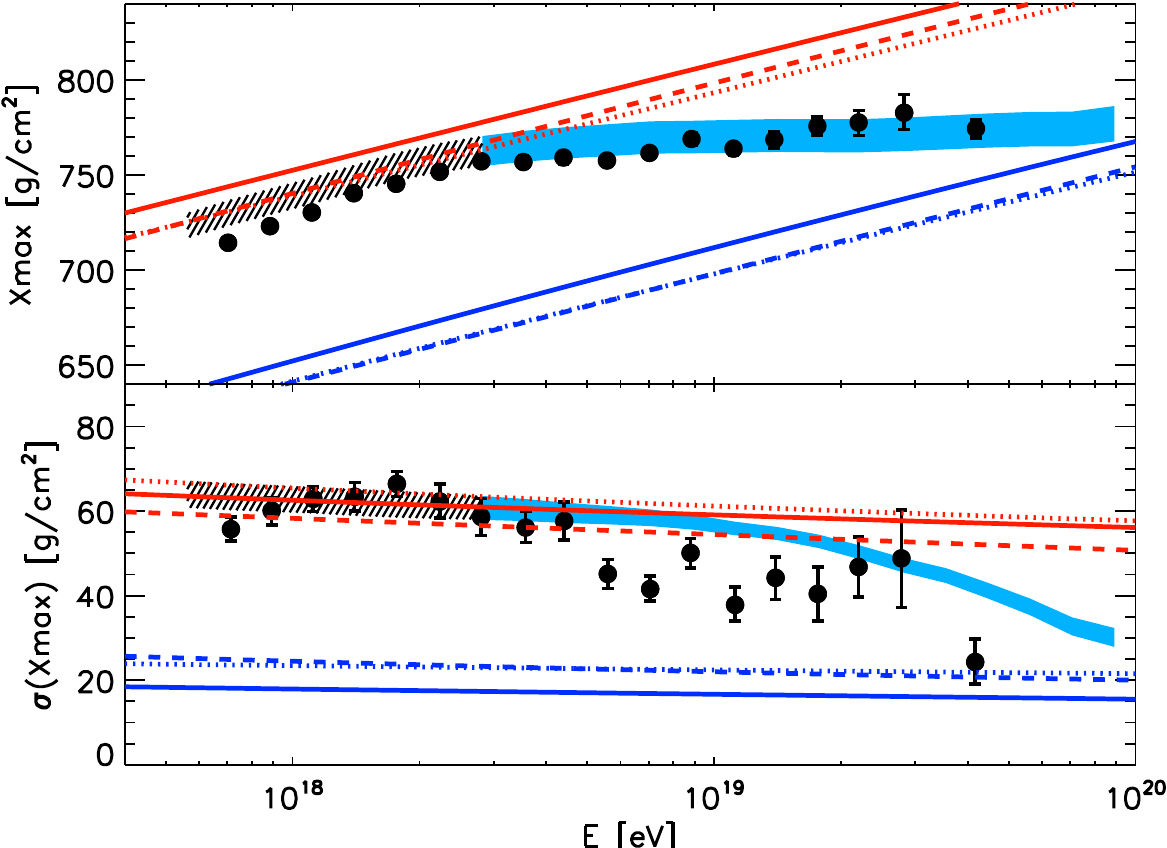,width=0.85\columnwidth,clip=}
\caption{\label{fig:cr} Up: Spectrum of UHECRs from newborn pulsars, assuming source emissivity following SFR and  injection composition: $50\%$ H,  $30\%$ CNO and $20\%$ Fe. Overlaid are measurements by the Auger Observatory \cite{ThePierreAuger:2013eja} and Telescope Array \cite{TAicrc11} with energy rescaling suggested in \cite{Dawson:2013wsa}. 
Bottom: values of estimations of UHECR composition, $\langle X_{\rm max} \rangle$ and RMS($X_{\rm max}$) of the Auger  data~\cite{ThePierreAuger:2013eja} (black crosses) and simulation results with pulsar sources  (blue shaded region where pulsars contribute to more than $80\%$ of the total flux, hashed region for less). Three hadronic interaction models, EPOS-LHC (solid), QGSJetII-04 (dotted) and Sibyll2.1 (dash) are used to estimate the range of $\langle X_{\rm max} \rangle$  and RMS($X_{\rm max}$) \cite{DeDomenico:2013wwa}.  The red and dark blue lines correspond to $100\%$ P and $100\%$ Fe. 
}
\end{figure}

Figure~\ref{fig:cr} shows the spectrum and composition of cosmic rays from extragalactic newborn pulsars for our best fit parameters to the Auger data (see also Ref.~\cite{FKO13}), assuming a source emissivity following the SFR, an ejecta mass $M_{\rm ej}=10\,M_\odot$ and wind acceleration efficiency $\eta=0.3$. The injected composition is $50\% $ H,  $30\% $ CNO,  $20\% $ Fe (injected protons can be mostly interchanged to Helium without affecting the spectrum significantly \citep{FKO13}). The overall normalization factor $f_{\rm s}=0.1$.

\begin{figure}[t]
\centering
\epsfig{file=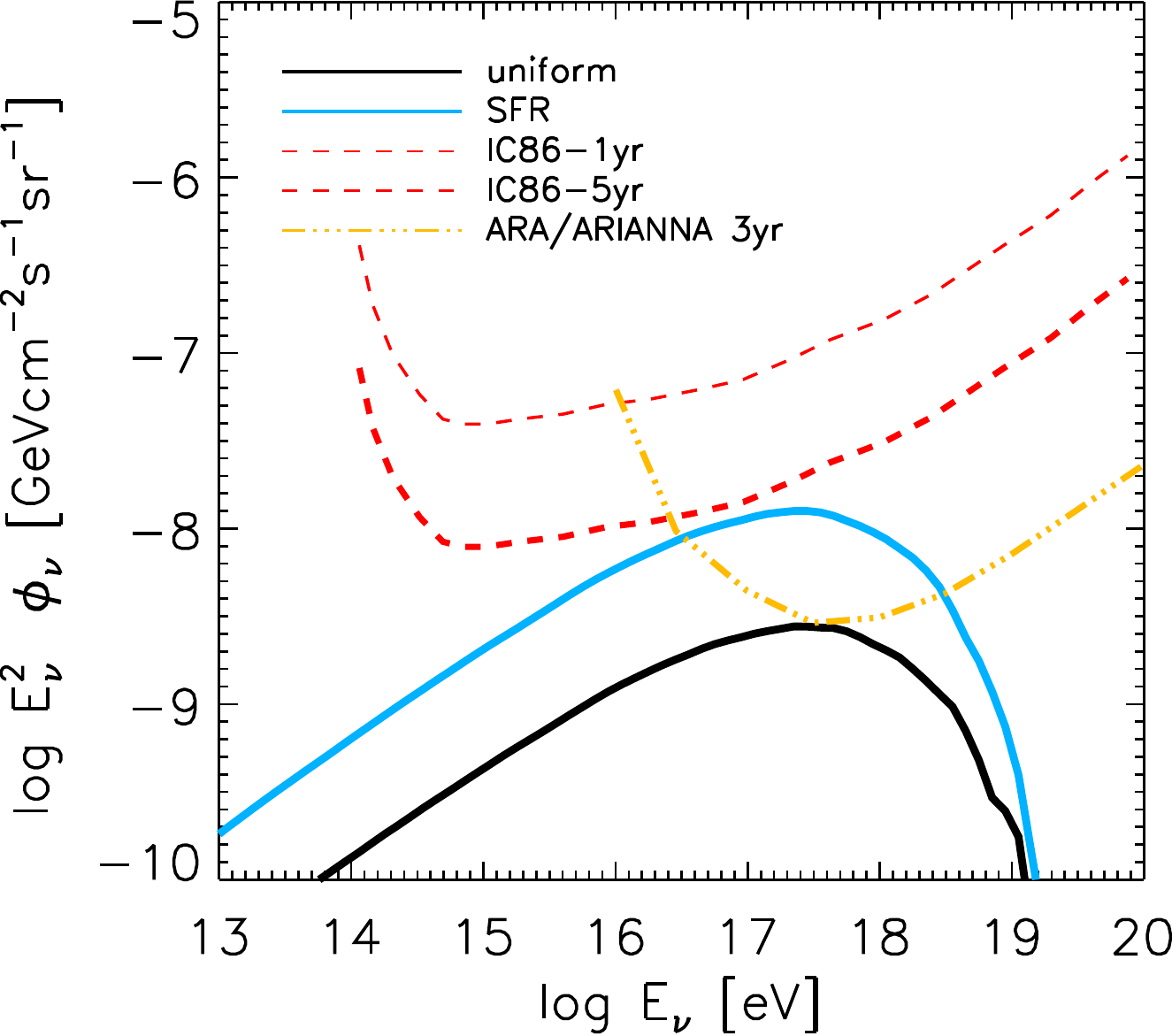,width=0.95\columnwidth,clip=}
\caption{\label{fig:neu} The diffuse neutrino flux ($\nu_\mu+\nu_e+\nu_\tau$ after neutrino mixing in space) from an extragalactic newborn pulsar population that would produce the measured UHECRs. The source emissivity follows the SFR (blue) or is uniform over time (black). Overlaid  are all flavor neutrino flux sensitivities of the IceCube detector after 1 year (red thin dash)  and 5 years (red thick dash) of operations \citep{Abbasi:2011zx}, and the expected 3 year ARA-37 sensitivities (orange dash dotted) \citep{Allison:2011wk}. }
\end{figure}

The associated diffuse neutrino fluxes are shown in Figure~\ref{fig:neu}. The case corresponding to the cosmic-ray counterpart shown in Fig.~\ref{fig:cr}, with SFR source emissivity evolution is drawn in blue. The black line represents the flux for a uniform source emissivity. The  flux is higher in the SFR case by a factor of $5$, which is the ratio between the total numbers of sources in these two cases. 

The neutrino spectrum consists of three components. Below $\sim 10^{16}\,\rm eV$, the spectrum can be described as a single power law with index $1.7$. This energy range corresponds to pulsars that spin relatively slowly with spin period $P\gtrsim 20\,\rm ms$. Only few interactions happen as the ejecta is mostly diluted when cosmic rays are produced. The neutrino spectrum hence roughly follows the cosmic ray spectrum, which is $E^{-1}$ at injection and softened to $E^{-1.7}$ due to the $(B, P)$ distribution. Between $10^{16}$ and $10^{18.8}\,\rm eV$, cosmic rays accelerated by the fast spinning pulsars undergo severe interactions with the baryons in the ejecta, resulting in an accumulation of neutrinos from secondary nuclei and pions that soften the spectrum to $E^{-2}$. Above $10^{18.8}\,\rm eV$, the spectrum cuts off as $P$ reaches its theoretically allowed minimum $P_{\rm min}=0.4\,\rm ms$ \citep{Haensel99} and $f(B)$ is small in the tail of the distribution.  Note that the neutrino spectrum has a peak at $\sim0.1-1$~EeV, implying nucleons with $\sim2-20$~EeV, and such UHECR nucleon production is possible in newborn magnetars \cite{Murase09}.  To be more conservative, we assume here that $f(P)$ cuts sharply at $P_{\rm min}$ instead of piling up, as was done in Ref.~\cite{FKO13}; the resulting difference is however negligible. 

The all flavor neutrino flux sensitivities of the IceCube detector after one year and five years of operation are shown in Figure~\ref{fig:neu} \citep{Abbasi:2011zx}, as well as the projected ARA-37 3-year sensitivity \citep{Allison:2011wk}.    The neutrino flux peak at around $10^{17.5}$ eV is found to be comparable to the IceCube 5-year sensitivity \citep{Abbasi:2011zx} in the SFR case, and to the 3-year ARA sensitivity  \citep{Allison:2011wk}  in the uniform case.  The flux is slightly smaller than the Waxman-Bahcall landmark and the prediction of the fast-spinning magnetar model. This is because efficient neutrino production occurs only in a fraction of fast-spinning pulsars and most pulsars can release cosmic-ray nuclei without depletion by meson production. 

The cosmogenic neutrinos produced during the intergalactic propagation are not shown in Figure~\ref{fig:neu}. This flux would be of the order of the SFR case with mixed composition in Ref.~\cite{KAO10}, represented by the lower boundary of the gray shaded region of their Fig.~9. The flux is below $\sim 6\times 10^{-9}\,\rm GeV\,cm^{-2}\,s^{-1}\,sr^{-1}$, and is subdominant compared to the neutrino contribution from the source region discussed here.

\section{Discussion}

The diffuse neutrino flux in the uniform case can be almost viewed as an unavoidable neutrino flux in the newborn pulsar scenario for UHECRs.  
As shown in Fig.~1, due to $f_{pp}>1$ at the time when $\sim Z{10}^{19}$~eV cosmic rays are accelerated, the pion production efficiency is the order of unity as long as $M_{\rm ej}\gtrsim3M_\odot$ (corresponding to $\beta\lesssim0.05$).  Also, the neutrino flux is insensitive to the injection composition because neutrinos are efficiently produced at relatively early times.  Ions are injected with a rate $\dot{N}=2\pi^2\,BR^3/P^2Zec$ and act effectively as $A$ nucleons in hadronic interactions (so that the energy of neutrinos from any species with mass number $A$ and charge $Z$ is proportional to $Z/A\sim0.5$). A minimum acceleration efficiency $\eta$ is a fitting subparameter, but our results on the neutrino flux does not change when $\eta\gtrsim0.1$ required for UHECR production.

{Our minimal pulsar scenario for UHECR predicts the diffuse neutrino flux of  $\sim\,3\times{10}^{-9}~{\rm GeV}~{\rm cm}^{-2}~{\rm s}^{-1}~{\rm sr}^{-1}$.  A lower neutrino flux than the one predicted in Fig.~\ref{fig:neu} is possible only by adding one of the following assumptions:  i) a jet puncture, expected only for high-power winds \cite{Murase09}, ii) ``shredding" of the envelope through Rayleigh-Taylor instabilities \citep{Arons03}, which could happen if $E_{\rm rot}>E_{\rm ej}$, iii) a thinner ejecta, for low-mass envelope or accretion-induced collapses. However, all the cases are nontypical and expected in rare types of supernovae.  Also, particles that would escape without interactions would not produce secondary lighter nuclei at lower energies, and it is not clear whether the produced cosmic rays can fit the observed composition and the spectrum.

In 2012, the IceCube Collaboration announced the first observation of two PeV neutrino-induced events during the combined IC-79/IC-86 data period \citep{Aartsen:2013bka}. A recent follow-up analysis of the same data found 26 additional events at lower energies \citep{Aartsen:2013pza}. No event has been observed yet at higher energies. 
Our model predicts a neutrino peak at $0.1-1\,\rm EeV$, and a flux about an order of magnitude lower than the observed flux around PeV energies (a softer injection spectrum would lead to fewer UHECR interactions and would not add much neutrino flux at this energy).  In principle, having a neutrino peak at PeV energies is possible for $\eta\ll0.1$, but the UHECR data cannot be explained at the same time.  Then, to account for the diffuse PeV neutrino flux, other possibilities should be invoked.  At present, there are various theories that are compatible with the IceCube data at PeV energies (e.g., Refs.~\cite{Kalashev:2013vba,Murase:2013ffa,Murase:2013rfa,Anchordoqui:2013qsi}), including pre-IceCube predictions (see Refs.~\cite{Murase:2013ffa,Murase:2013rfa} and references therein).

One of the caveats in the newborn pulsar scenario for UHECR is uncertainty in particle acceleration mechanisms. Though the viability of this scenario depends on pair-loading in the equatorial wind and acceleration processes, since the cosmic-ray flux is normalized by the UHECR observations, the diffuse neutrino flux we predict in the EeV range does not depend on the underlying details.  Note that Fermi mechanisms lead to softer cosmic-ray injections than the hard $E^{-1}$-spectrum, but the spectrum after escape from the ejecta is almost the same.  The secondary products from interactions with the ejecta soften the spectrum to $E^{-2}$ above $10^{17}\,\rm eV$, and this effect is less pronounced in the case of a softer intrinsic spectrum, because less high energy primaries are injected. The combination of these antagonist effects argues also against a significant change in the neutrino flux between $0.1-1\,$EeV, for softer injections.

Another possible issue is photodisintegration due to interactions with ambient photons.  As already noted above, thermal and nonthermal radiations are also expected to lead to photodisintegration \cite{Murase09,FKO12,KPO13}. Ref. \cite{Murase09} showed that the thermal radiation background over the supernova ejecta can play a role in the magnetar case. In addition, X-ray and gamma-ray nonthermal fields in the pulsar wind nebula could be strong enough to compete with the hadronic channel \cite{mdt14}, where our neutrino predictions can then be relatively conservative.  Note however that if photohadronic neutrinos are dominant, nuclei would be mostly disintegrated due to the larger photodisintegration cross sections \cite{Murase:2010gj}, and the scenario would fail at satisfying our primary requirement of reproducing the Auger data.

Before we end this section, we comment on how we can test the newborn pulsar origin.  As shown in this paper, measurements of the diffuse neutrino flux is very powerful in the sense that nondetection can strongly constrain the scenario. However, if diffuse neutrinos are detected, it becomes important to discriminate this possibility from the other scenarios.  First, a single source detection is difficult but not impossible.  At high energies, the atmospheric neutrino background is essentially negligible, so it is possible to identify a single source up to a few Mpc \cite{Murase09}. Furthermore, pulsars allowing UHECR acceleration have to be fast-spinning, so that the rotation energy can affect supernova dynamics. Thus, neutrino events should be associated with luminous or energetic supernovae powered by pulsars~\cite{2004ApJ...611..380T,Komissarov07,2010ApJ...717..245K,2010ApJ...719L.204W,KPO13}, so stacking such bright supernovae within dozens of Mpc would also be useful \cite{Murase09}. Second, in this scenario, not only neutrinos but also hadronic gamma rays should be produced. In the late phase, emission of cascaded GeV-TeV gamma rays is unavoidable, which may be detected by ground-based gamma-ray detectors. In addition, if target photon fields are thermal, even ultrahigh energy gamma rays may escape, which provides a useful probe of UHECR accelerators within dozens of Mpc \cite{mur09}.  All the details of gamma rays signatures are beyond the scope of this paper, which are left for future work. 
Third, UHECR measurements are useful for consistency checks, although source identification is very difficult when UHECR sources are transients and the composition is heavy.  Newborn pulsars should be regarded as transient UHECR sources. This is because the emission duration of UHECRs from a single pulsar would be in a scale up to years, much less than the delay caused by the particles' deflection in the extragalactic magnetic field~\citep{FKO12,2012ApJ...748....9T}. This is even the case if UHECRs are largely heavy nuclei, since the Galactic magnetic field also causes significant time delays for nuclei.  Moreover, as our pervious work suggested~\citep{FKO13}, the percentage of the pulsar population that are capable to accelerate particles to above 10 EeV is just about 0.3\%. Thus, the transient nature and the rareness of such sources can significantly decrease the anisotropy from light nuclei from a potential nearby source, especially if the extragalactic magnetic field is relatively strong.  The anisotropy signal is significantly diminished for nuclei since the deflection angle at the same energy is proportional to the inverse of atomic number.  Therefore, no particularly striking anisotropy features are expected with the current Auger statistics, even though future generation detectors could detect some anisotropy signal (see Refs.~\cite{2012APh....35..767T,Rouille14}).  
Note that turbulent Galactic magnetic fields are strong enough to diminish strong anisotropy signals~\cite{2011APh....35..192G}, and they are further smeared out with extragalactic magnetic fields in local structured regions~\citep{1999PhRvD..59b3001B}.

\section{Summary}

We have shown that a newborn pulsar scenario that explains the UHECR data necessarily leads to efficient neutrino production.  For the plausible source evolution, the diffuse neutrino flux lies sensibly  between the IceCube-5-yr and ARA-3-yr sensitivities in the ${10}^{18}$~eV energy range. The newborn pulsar scenario has a strong prediction for the diffuse neutrino flux in the sense that nondetections of neutrinos at these energies in the next decade will rule out the minimal pulsar scenario.  
Successful detections of the diffuse neutrino flux would not necessarily mean the confirmation of the pulsar scenario. To establish the newborn pulsar scenario for UHECR, more dedicated multimessenger searches are needed but they could provide us with a unique opportunity of studying ion acceleration in newborn pulsars.

\acknowledgements
We thank T. Pierog for his help with the hadronic interaction code EPOS and the Auger group at the University of Chicago for very fruitful discussions. KK thanks KICP for its kind support and hospitality. This work was supported by the NSF grant PHY-1068696 at  the University of Chicago, and the Kavli Institute for Cosmological Physics through grant NSF PHY-1125897 and an endowment from the Kavli Foundation. KK and AVO acknowledge support from PNHE.  KF and AVO acknowledge financial support from NASA 11-APRA-0066. KM is supported by NASA through Hubble Fellowship, Grant No. 51310.01 awarded by the STScI, which is operated by the Association of Universities for Research in Astronomy, Inc., for NASA, under Contract No. NAS 5-26555.

\bibliography{FKMO14}

\end{document}